# Robust GPS – SMS Communication Channel for the AVL System

Joseph Skobla, Andrew Young

The University of West Indies, Physics Department, Kingston, Jamaica

Universal Preprocessing GPS SMS Communication Unit (UPCU) was developed as a part of the UWI microtracking system. A GSM cellular Short Messaging Service is the main method of delivering tracking information to the central base station.

The information includes asset ID, longitude, latitude, altitude, speed and direction as well as the time the message was sent. The design of the unit is implemented as a bi-directional SMS system. Message can be originated either at the mobile unit or at the base station. The base station can send a SMS message requesting information from the Mobile unit, the unit can also be configured for a different mode of operation from the base station. A set of special configuration messages was developed.

The hardware of the mobile unit is made up of three main subsections: the GPS receiver, a GSM modem and a preprocessing unit. The preprocessing unit extracts specific information from the GPS receiver and uses this it to reconstruct a new message with all the necessary asset information. This new message is sent as part of a SMS text to the base station using the GSM modem. The base station can send configuration or data request messages to the mobile unit. This data is received through the GSM modem. Messages receive via the modem are not stored in the SIM memory of the module but is sent directly to the serial port of the unit.

The UPCU is a micro-controller based buffering and message extraction system. It has two RS232 ports, one is used to receive and send data to the modem and the other is used to receive the data from the GPS receiver. The preprocessing unit has a built-in timer which is capable of sending navigational information to the base station on each timeout of the timer. A SMS message is also sent to the base station when the status report button is pressed on the mobile unit.

Tracking information is sent to the central base station where it is processed. The UPCU was designed for targeting applications that require tracking information on a periodic basis rather than establishing a connection and sending the information continuously. The unit is capable of sending information to the base station at a rate of approximetly one message per second across a standard data communication channel.

The software can be divided into two sections: the base station interface software and the processing unit system software. The base station interface software is used to convert the data received at the base station into some format that is compatible with a specific application software that uses the data, example mapping or data base software.

The preprocessing unit system software is the program running on the micro-controller. The system controls all the interrupt resources, it is also responsible for buffer and RS232 resource management. SMS messages support a maximum of 160 characters.

The nature of SMS messages and its universal implementation across GSM ( or by another cellular network)  it is ideal for transmitting messages with tracking information.

The SMS messages can deliver all the major information needed to provide a tracking service not for only vehicles, but unit can be used a component of any remote sensing application with a programmable mode of operation.